\documentclass[journal]{IEEEtran}
\usepackage{algorithm}
\usepackage{algorithmic}
\usepackage{cite}
\usepackage{bbm}
\usepackage{amsmath}
\usepackage{amsfonts,amssymb}
\usepackage{graphicx}
\allowdisplaybreaks[4]
\usepackage{enumitem}
\usepackage{booktabs}
\usepackage{xcolor,colortbl}

 \usepackage{bm}
 \usepackage{multirow}
\usepackage{amsthm}
\usepackage{amssymb}
\usepackage{pgfplots}
\usepackage[a]{esvect}

\usepackage{comment}
\definecolor{myblue}{RGB}{105,89,205}
\definecolor{myorange}{RGB}{238,92,66}
\definecolor{mygrey}{RGB}{205,201,201}
\definecolor{myred}{RGB}{205 38 38}

\begin{document}

\title{Model-Free Optimal Control of Inverter for Dynamic Voltage Support}
\author{
	Yifei~Guo,~\IEEEmembership{Member,~IEEE,}
	Bikash C. Pal,~\IEEEmembership{Fellow,~IEEE,}
	Rabih A. Jabr,~\IEEEmembership{Fellow,~IEEE}
\thanks{This work was supported by the Resilient Operation of Sustainable Energy
Systems (ROSES) UK-China (EPSRC-NSFC) Programme on Sustainable Energy Supply (EP/T021713/1, NSFC-52061635102). (\emph{Corresponding author: Bikash C. Pal})}
\thanks{Yifei Guo and Bikash C. Pal are with the Electrical and Electronic
Engineering Department, Imperial College London, London SW7 2AZ, U.K. (e-mail: yifei.guo@imperial.ac.uk; b.pal@imperial.ac.uk)}
\thanks{Rabih A. Jabr is with the Department of Electrical and Computer
Engineering, American University of Beirut, Beirut 1107 2020, Lebanon
(e-mail: rabih.jabr@aub.edu.lb)}
}
\markboth{}
{Shell \MakeLowercase{\textit{et al.}}: Bare Demo of IEEEtran.cls for IEEE Journals}
\maketitle

\begin{abstract} Inverter-based resources (IBRs) are required to provide dynamic voltage support (DVS) during voltage dips to enhance the low-voltage ride-through capability. In this paper, we develop a model-free control method to achieve the optimal DVS (ODVS) without relying on the knowledge of grid parameters. Delving into the optimum trajectory of the ODVS problem, it is found that either the current constraint and the maximum active power constraint of IBRs are binding or one of the constraints is binding. This inspires us to search for the optimum in a closed-loop way by a perturb-and-observe (P\&O)-based optimum seeking (OS) controller with either the power factor angle or the reactive current being the manipulated (perturbed) variable. The system is guaranteed to converge asymptotically to the optimum provided the stepsize sequence is diminishing and non-summable. 
The proposed model-free optimal control is finally implemented within a single-stage photovoltaic (PV) system, where dynamic simulations demonstrate the optimal and fast DVS performance.
\end{abstract}

\section{Introduction}
\IEEEPARstart{T}HERE has been a broad consensus among worldwide grid codes on the operational requirements  for inverter-based resources (IBRs) in response to voltage disturbances---IBRs should provide dynamic voltage support (DVS); for example, a certain amount of reactive current provision in proportion to the retained voltage at the point of connection (PoC) is usually mandatory during grid voltage dips \cite{IEEE1547,TH2019,EDK2015,VDE2017,GB19963,GB19964,AS2019}.

Although there is little doubt that the reactive current injection can help support the voltage, prioritizing reactive current or even purely injecting reactive current does not maximize the DVS. Furthermore, with  the increasing integration of IBRs and the retirement of synchronous generators in power systems, IBRs play an increasingly vital role in improving system immunity against disturbances. So, it is imperative to explore a better way for IBRs to provide DVS, as simple heuristic rules usually cannot unlock the maximum DVS capability of IBRs and may even cause instability \cite{GO2014,WB2015}.

Recent years have seen a tremendous surge of interest in developing advanced DVS strategies of IBRs based on optimization, specifically for boosting the positive-sequence voltage magnitude \cite{KK2017,CA2016,CA2018,SMA2019,naidu2021energy,YG2021}. The work in \cite{KK2017} formulates an optimization model for maximizing the PoC voltage where the maximum fault current injection is imposed, based on a sensitivity-based linearized relationship between the PoC voltage  and active and reactive currents of IBRs. Rather than relying on the linear approximation, the studies in \cite{CA2016,CA2018,SMA2019} present the optimization models for maximizing the PoC voltage subject to the maximum current limit, based on the entire nonlinear voltage-current relationship. The work in \cite{naidu2021energy} further takes into account a load impedance at the PoC. A thorough  optimality analysis of the optimal DVS (ODVS) problem is provided in \cite{YG2021}, considering the current, active power, and synchronization stability constraints of IBRs.

It has been extensively verified in previous studies \cite{KK2017,CA2016,CA2018,SMA2019,naidu2021energy,YG2021} that the current injection of IBRs that yields the maximum (optimal) DVS depends not only on the current and power limits of IBRs, but also on the Th{\'e}venin equivalent of the grid seen from the PoC; the underlying mathematical relationship has been revealed  \cite{YG2021}. Most of those studies assume the availability of the grid model (grid voltage and/or impedance). Indeed, it is straightforward to implement the ODVS based on the knowledge of grid parameters. However, as there are typically stringent requirements on the response time of DVS\footnote{For example, solar and wind farms shall provide DVS within 30 and 60 ms since a voltage dip occurs, respectively, in China \cite{GB19963,GB19964}.}, the time window available for grid estimation may be too short even for most state-of-the-art techniques \cite{abdi2021pmu,sobhy2022online,zonetti2022observer,mohammed2020online} to work effectively and reliably, especially during dramatic voltage disturbances\footnote{For example, the online event-based grid impedance estimation technique for IBRs requires 750 ms to update the result \cite{mohammed2020online}.}. 
This reveals the fact that it is very challenging  in practice to achieve the ODVS based on the full knowledge of grid parameters; note that using erroneous grid parameters could not only result in loss of optimality, but may even  cause instability, e.g., loss of synchronism (LoS) or power imbalance between ac and dc sides. The authors in \cite{KK2017} realize the difficulty in computing the sensitivity coefficients without knowing external grid conditions and thus propose a compromise solution: outputting the maximum and equal active and reactive currents regardless of grid conditions. Such a strategy eliminates the need for grid estimation; however, it may not only sacrifice the optimality but also induce instability under certain conditions. 

To overcome these challenges, this work explores the ODVS control of IBRs that does not rely on a feedforward grid parameter estimation. Without the availability of grid parameters, the ODVS becomes a gray-box optimization problem\footnote{Note that the model structure is imposed, i.e., the Th{\'e}venin equivalent still holds, but the model parameters are no longer assumed to be known.}. 
In this context, we propose a model-free control method to seek the optimum of a gray-box ODVS problem in an online fashion. Revisiting the optimum trajectory of the ODVS problem derived in our earlier work \cite{YG2021}, it is found that the current constraint (CC) and/or the maximum active power constraint (MAPC) of the IBR is binding; that is, the IBR should either output the maximum current or maximum active power to achieve ODVS. This indicates that one can search for the optimum along the feasible space boundary on which the optimum is instead of exhaustively searching over the whole feasible space. To do so, we first discover the functional relationship between the PoC voltage and power factor angle (reactive current) based on the condition that the IBR operates along the boundary of CC (MAPC). It is proven that the functions have a unique local maximum within a specific domain, which exactly corresponds to the optimum of the ODVS problem. In this context, solving the two-dimensional (2-D) ODVS 
problem can be reduced to a one-dimensional (1-D) search. This inspires a perturb-and-observe (P\&O)-based optimum seeking (OS) with either the power factor angle or the reactive current being the manipulated (perturbed) variable. 
The criteria associated with the stepsize rule design are given, guaranteeing that the P\&O-based OS can converge to the optimum without steady-state oscillations. Finally, we detail the implementation method of the model-free optimal DVS control and embed  it into a single-stage photovoltaic (PV) generation system. Its performance is tested and compared with other existing practices under various conditions. 

The rest of this paper is organized as follows. Section II revisits the problem statement and the optimum trajectory. Section III introduces the dimensionality reduction approach. In Section IV, we develop the P\&O-based OS algorithm. Section V presents the implementation method of the model-free DVS control embedded within a single-stage PV generation system. In Section VI, dynamic simulations are conducted, followed by conclusions. The proofs are collected in the Appendix.

\section{Preliminaries}
\subsection{Problem Formulation}
\begin{figure}[t]
    \centering
    \includegraphics[width=3.5in]{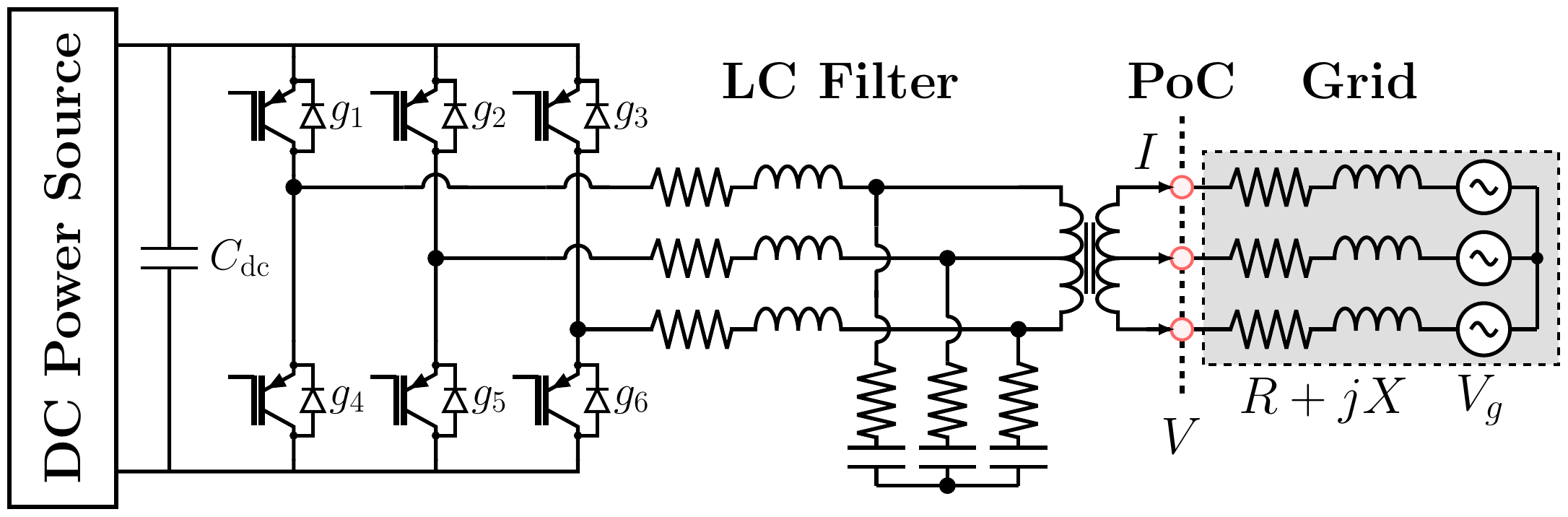}
    \caption{Schematic diagram of a three-phase grid-connected IBR. }
    \label{3phIBR}
\end{figure}
The schematic diagram of a three-phase grid-connected IBR is shown in Fig. \ref{3phIBR}, where $V_g$ and $R+jX$ are the positive-sequence grid voltage and impedance. The three-phase inverter employs a typical cascaded control loops, i.e., an outer loop that regulates the power/voltage, and an inner loop that controls the current. The standard dual current control strategy under the dq-frame that allows the decoupled control of positive-and negative-sequence currents is adopted. The system modeling has been detailed in \cite{YG2021}, so is omitted here.\footnote{Since we focus only on the positive-sequence voltage support in this work, the negative-sequence current is kept to zero. But note that advanced negative-sequence control is also compatible, as long as it can be well coordinated with the positive-sequence control to respect all the physical limits.}

To reduce the risk of being tripped by the under-voltage protection, IBRs shall provide DVS during voltage dips. So, the ODVS seeks to maximize the positive-sequence PoC voltage magnitude ($V$) by optimizing the IBR's active and reactive current injections ($I_{\rm d}$ and $I_{\rm q}$) while respecting all the physical constraints. It is formulated as
\begin{subequations}\label{ODVS}
\begin{align}
\underset{I_{\rm d},I_{\rm q}}{\rm maximize}\hspace{3mm} &V(I_{\rm d},I_{{\rm q}})\\
{\rm subject\,to}\hspace{3mm} & g_c(I_{\rm d},I_{\rm q})\leq 0\hspace{5mm}{\rm Current\, Constraint}\\
&g_{p}(I_{\rm d},I_{\rm q})\leq 0\hspace{5mm}{\rm Max.\, Power\, Constraint}
\end{align}
\end{subequations}
where 
\begin{align}
    V(I_{\rm d},I_{{\rm q}})&:=\sqrt{V_g^2-(RI_{\rm q}+XI_{\rm d})^2}+RI_{\rm d}-XI_{\rm q}\\
    g_c(I_{\rm d},I_{\rm q})&:=I_{\rm d}^2+I_{\rm q}^2-I_{\rm max}^2\\
    g_{p}(I_{\rm d},I_{\rm q})&:=\frac{3}{2}V(I_{\rm d},I_{\rm q})I_{\rm d}-P_{\rm max}.
\end{align}

$I_{\rm max}>0$ and $P_{\rm max}>0$ denote the IBR's maximum current limit and the maximum power limit. The maximum available power $P_{\rm max}$ depends on the dc-side energy source, e.g., wind, solar, battery storage, etc., and may thus vary over a wide range.  Note that, the synchronization stability constraint (SSC) 
\begin{align}
    g_{s}(I_{\rm d},I_{\rm q})&:=|RI_{\rm q}+XI_{\rm d}|-V_g\leq0
\end{align}
is hidden in (\ref{ODVS}). If this constraint is violated, the IBR will lose synchronism with the grid.

Compared with the original formulation presented in \cite{YG2021}, we relax all the constraints that are definitely non-binding (inactive)  under different model parameters ($V_g, R, X, I_{\rm max}$ and $P_{\rm max}$). Clearly, this does not affect the solution.
\subsection{Optimum Trajectory}
If all the model parameters ($V_g,R,X,I_{\rm max},$ and $P_{\rm max}$) are available, the ODVS problem (\ref{ODVS}) can easily be solved according to the optimality analysis conducted in \cite{YG2021}. Not surprisingly, the optimum  is strongly related to the model parameters. Let $P_{\rm max}$ decrease from a sufficiently large value  to zero and fix other parameters. The optimum of (\ref{ODVS}) may experience three stages: {$S_1\rightarrow S_2\rightarrow S_3$}, and correspondingly, there exists two boundary conditions, {C1} and {C2}, that determine which stage is valid, which is detailed as follows:
\begin{itemize}
    \item Stage 1 (if C1 holds):
     \begin{align}\label{OPTS1}
               S_1:  \left\{\hspace{-2mm}  \begin{array}{l}
        I_{\rm d}^\star=\dfrac{R}{Z}{I}_{\rm max},\\[2mm]
         I_{\rm q}^\star=-\dfrac{X}{Z}{I}_{\rm max}
             \end{array}\right.   
            \end{align}
            where $Z:=\sqrt{R^2+X^2}$.
      \item Stage 2 (if neither C1 nor C2 holds): $ S_2$  is the unique solution satisfying the following conditions:
\begin{align}\label{OPTS2}
     S_2:   \left\{\hspace{-2mm}  \begin{array}{l}
        g_c( I_{\rm d}^\star, I_{\rm q}^\star)=0\\[2mm]
         g_{p}( I_{\rm d}^\star, I_{\rm q}^\star)=0\\[2mm]
         I_{\rm d}^\star\geq0,  I_{\rm q}^\star\leq0.
             \end{array}\right.   
    \end{align}
\item Stage 3 (if C2 holds): 
    \begin{align}\label{OPTS3}
       \hspace{-3mm} S_3:  \left\{\hspace{-2mm}
         \begin{array}{l}
I_{\rm d}^\star=-\dfrac{1}{2Z}\left(V_g-\sqrt{V_g^2+\dfrac{8R{P}_{\rm max}}{3}}\right)\\[5mm]
         I_{\rm q}^\star=-\dfrac{X}{2RZ}\left(V_g+\sqrt{V_g^2+\dfrac{8R{P}_{\rm max}}{3}}\right).     
         \end{array}
         \right.
    \end{align}
\end{itemize}

The boundary conditions C1 and C2 are
   \begin{align}
    \label{C1}
   {\rm C1:}&\hspace{1mm}  {P}_{\rm max}\geq P_b\\
    {\rm C2}:&\hspace{1mm} I_{\rm max}\geq I_b
   \end{align}
where $P_b$ and $I_b$ are two specific parameters associated with the grid parameters.

\begin{figure}[t]
    \centering
    \includegraphics[width=3.5in]{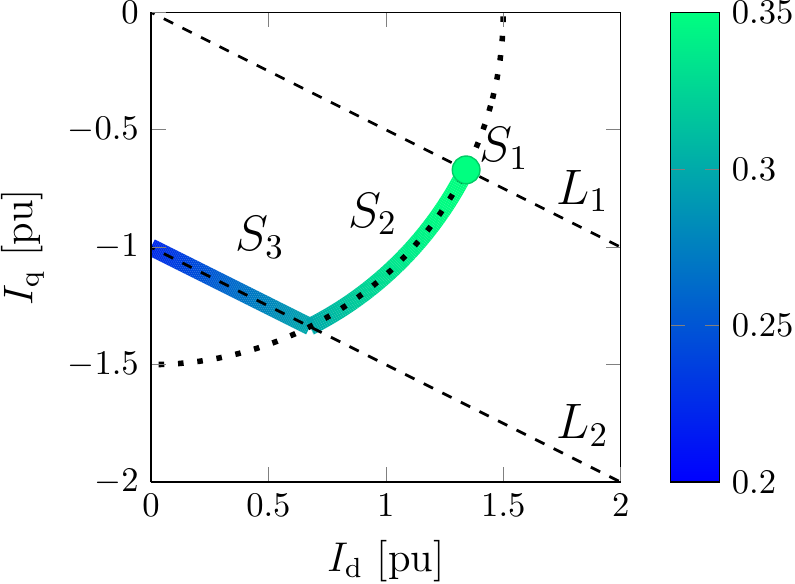}
    \caption{Optimum trajectory of optimal DVS problem under varying $P_{\rm max}$. }
    \label{OptTraj}
\end{figure}

The optimum trajectory under varying $P_{\rm max}$ is illustrated in Fig. \ref{OptTraj}, where the color bar represents the resultant maximum PoC voltage; the dotted arc is the maximum current boundary; ${L}_1$ and ${L}_2$ are two parallel straight lines: $RI_{\rm q}+XI_{\rm d}=0$ and $RI_{\rm q}+XI_{\rm d}+V_gX/Z=0$, respectively. In geometry, $S_1$ is the intersection  of the maximum current boundary and the straight line ${L}_1$, $S_2$ is the intersection  of the maximum current boundary and the maximum power boundary, and $S_3$ is the intersection of the maximum power boundary and the straight line ${L}_2$. 
Keep in mind that if the grid is strong, and/or the voltage dip is mild (so that $L_2$ does not intersect with the maximum current boundary in the fourth quadrant), $S_3$ may not exist no matter how $P_{\rm max}$ varies. More details of formulation and optimality analysis can be referred to \cite{YG2021}.
\subsection{Gray-Box Optimization }
It is clear that the optimum is strongly related to the grid parameters ($R,X$, and $V_g$). However, as mentioned earlier, accurate grid parameters are often unavailable for DVS in practice, indicating that the above analytical solution cannot be straightforwardly used. In this context, optimization problem (\ref{ODVS}) becomes a gray-box optimization problem since some model parameters are unknown.
Solving such a gray-box optimization program is typically very challenging for many conventional numerical algorithms, in the sense that 1) the gradient information is unavailable, 2) the objective function cannot be directly evaluated, and 3) the constraint violations are hard to foresee. In addition, solution efficiency is also a big concern, as the DVS from IBRs should be fast and agile to comply with modern grid codes. So, in what follows, we will initiate a new way of solving the gray-box counterpart of (\ref{ODVS}).



\section{Dimensionality Reduction}
No gradient information makes it nontrivial to  solve for the optimum efficiently, in the sense that the best search direction among all the candidates is unknown. However, one exception could be solving a one-dimensional problem, where there are only two potential search directions, ``positive ($+1$)" and ``negative ($-1$)". This inspires us to explore a dimensionality reduction of the search, which provides the foundation for the later OS.
\subsection{Reduced Solution Space}
Looking into the optimum trajectory again, it can be found that CC is \emph{binding} if C1 holds, i.e., $g_c(I_{\rm d}^\star,I_{\rm q}^\star)=0$; MAPC is \emph{binding} if C1 does not hold, i.e., $g_{p}(I_{\rm d}^\star,I_{\rm q}^\star)=0$. This implies that one can search for the optimum either along the maximum current boundary
\begin{align}
    B_c:= \left\{(I_{\rm d},I_{\rm q}) : g_c(I_{\rm d},I_{\rm q})=0\right\},
\end{align}
 or along the maximum power boundary
 \begin{align}
    B_p:= \left\{(I_{\rm d},I_{\rm q}) : g_p(I_{\rm d},I_{\rm q})=0\right\},
\end{align}
instead of exhaustively searching over the whole 2-D feasible domain. Furthermore, over  ${B}_c$ or ${B}_p$, $I_{\rm d}$ and $I_{\rm q}$ are no longer mutually independent, enabling the dimensionality reduction.

\subsection{Dimensionality Reduction Based on $B_c$}
Let us introduce the power factor angle $\phi:={\rm atan2}(I_{\rm q},I_{\rm d})$. Then, we have $B_c=\big\{(I_{\rm max}\cos\phi,I_{\rm max}\sin\phi)|-180^\circ\leq\phi\leq180^\circ\big\}$. 
Correspondingly, the objective function $V$ over $B_c$ can be re-expressed as a function of $\phi$, i.e.,\footnote{For simplicity of exposition, we abuse the notation of $V(\cdot)$ depending on the context. This also applies to $g_c(\cdot)$, $g_p(\cdot)$, and $g_s(\cdot)$.}
\begin{align}
   V(\phi)=V(I_{\rm max}\cos\phi,I_{\rm max}\sin\phi).
\end{align}

We find that if C1 holds, $V$  strictly increases over $[\underline{\phi},\phi^\star]$ and strictly decreases over $[\phi^\star,\overline{\phi}]$, where 
\begin{align}\label{phimaxmin}
\phi^\star&:={\rm atan2}(-X,R)\\
[\underline{\phi},\overline{\phi}]&:=\left\{\phi\in[\underline{\phi_s},\overline{\phi_s}]: g_p(\varphi)\leq0, \forall\varphi\in[\underline{\phi_s},\phi]\right\}\\
[\underline{\phi_s},\overline{\phi_s}]&:=\left\{\phi\in[-90^\circ,0^\circ]: g_s(\phi)\leq0\right\}.
\end{align}

\begin{figure}[t]
    \centering
    \includegraphics[width=3.5in]{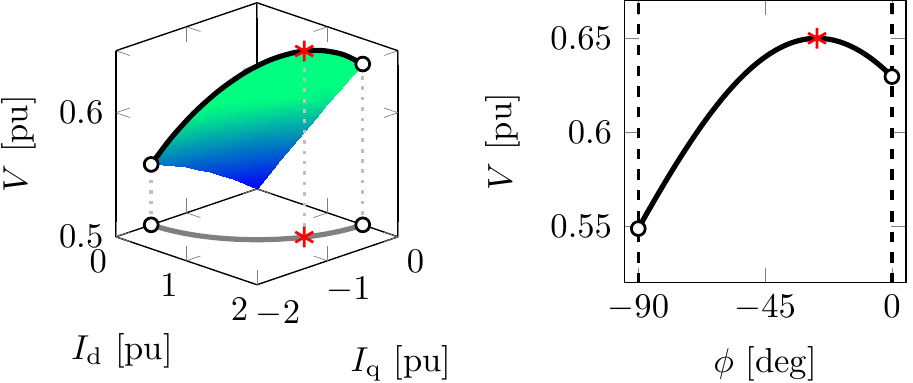}
    \includegraphics[width=3.5in]{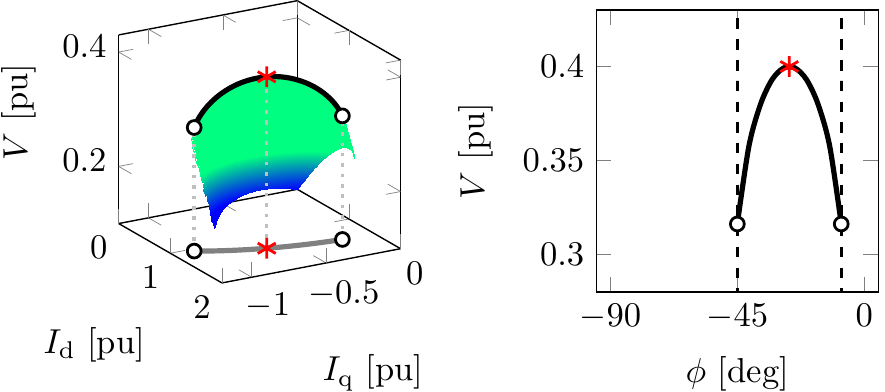}
    \caption{Dimensionality reduction based on $B_c$. Top: $V_g=0.5$ pu, $Z=0.1$ pu, $R/X=2$; bottom: $V_g=0.1$ pu, $Z=0.2$ pu, $R/X=2$.  }
    \label{3Dto2D_S1}
\end{figure}

This suggests that there exists an interval $[\underline{\phi},\overline{\phi}]$, over which $V$ has a unique local  maximizer $\phi^\star$, which exactly corresponds to $S_1$. So, solving (\ref{ODVS}) can be achieved by searching for $\phi^\star$ over $[\underline{\phi},\overline{\phi}]$ if C1 holds. Fig. \ref{3Dto2D_S1} gives two illustrative examples of dimensionality reduction under different grid conditions where we assume that $P_{\rm max}$ is sufficiently large. It shows that if the voltage dip is mild and/or the grid is strong, $[\underline{\phi},\overline{\phi}]$ will be  $[-90^\circ,0^\circ]$ (see the top example); if the voltage dip is sufficient and/or the grid is  weak,  $[\underline{\phi},\overline{\phi}]$ may shrink subject to SSC (see the bottom example where $\underline{\phi}=-45^\circ$, and $\overline{\phi}=-8.13^\circ$). If $P_{\rm max}$ takes a value such that $P_{\rm max}\geq P_b$ and  $g_p(\phi)>0$ for some $\phi\in[\underline{\phi_s},\overline{\phi_s}]$, $\Phi$ will further shrink subject to MAPC.
\subsection{Dimensionality Reduction Based on $B_p$}
For the solutions $(I_{\rm d},I_{\rm q})\in B_p$, there exists an implicit functional relationship between $I_{\rm d}$ and $I_{\rm q}$, i.e., $I_{\rm d}=\Psi(I_{\rm q})$ or $I_{\rm q}=\Psi^{-1}(I_{\rm d})$; consequently, the objective function $V$ over $B_p$ can be reduced to a function of solely $I_{\rm q}$, that is,
\begin{align}
   V(I_{\rm q}):=V\left(\Psi(I_{\rm q}),I_{\rm q}\right).
\end{align}

Then, it is found that if neither C1 nor C2 holds,  $V$  strictly decreases over $[\underline{I_{\rm q}},\overline{I_{\rm q}}]$, where the upper bound is 
\begin{align}\label{Iqmax}
    \overline{I_{\rm q}}=\dfrac{X}{2Z^2}\left(V_g-\sqrt{V_g^2+\frac{8Z^2P_{\rm max}}{3R}}\right)
\end{align}
and the lower bound $\underline{I_{\rm q}}$ uniquely satisfies
\begin{align}\label{Iqmin1}
     \left\{\hspace{-2mm}  \begin{array}{l}
         g_{p}\Big(\sqrt{I^2_{\rm max}-\underline{I_{\rm q}^2}},\underline{I_{\rm q}}\Big)=0\\[2mm]
         \underline{I_{\rm q}}\leq0.
             \end{array}\right.   
    \end{align}
It is not hard to see that $\underline{I_{\rm q}}$  is also the unique local maximizer $I_{\rm q}^\star$ over $[\underline{I_{\rm q}},\overline{I_{\rm q}}]$, which exactly corresponds to $S_2$.

If C2 holds, $V$ is strictly increasing over $[\underline{I_{\rm q}},I_{\rm q}^\star]$ and strictly decreasing over $[I_{\rm q}^\star,\overline{I_{\rm q}}]$, where 
$I_{\rm q}^\star$ is the same as  in (\ref{OPTS3}),  i.e., $S_3$; $\overline{I_{\rm q}}$ is the same as in (\ref{Iqmax}); $\underline{I_{\rm q}}$ is the same as  in (\ref{Iqmin1}) if 
\begin{align}\label{C3}
I_{\rm max}^2\geq\frac{R^2+Z^2}{2R^2Z^2}V_g^2+\frac{2P_{\rm max}}{3R}+\frac{X^2V_g}{2R^2Z^2}\sqrt{V_g^2+\frac{8Z^2RP_{\rm max}}{3X^2}}
\end{align}
otherwise,
\begin{align}\label{Iqmin2}
    \underline{I_{\rm q}}=-\frac{R^2+Z^2}{2RZ^2}V_g-\frac{X^2}{2RZ^2}\sqrt{V_g^2+\frac{8Z^2RP_{\rm max}}{3X^2}}.
\end{align}

\begin{figure}[t]
    \centering
    \includegraphics[width=3.5in]{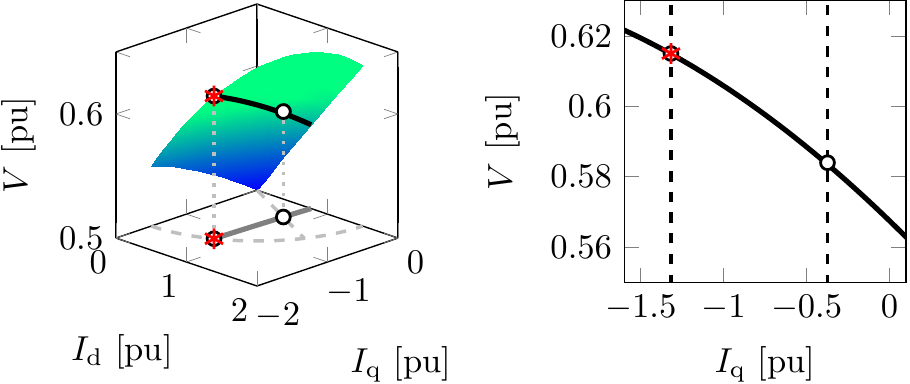}
    \includegraphics[width=3.5in]{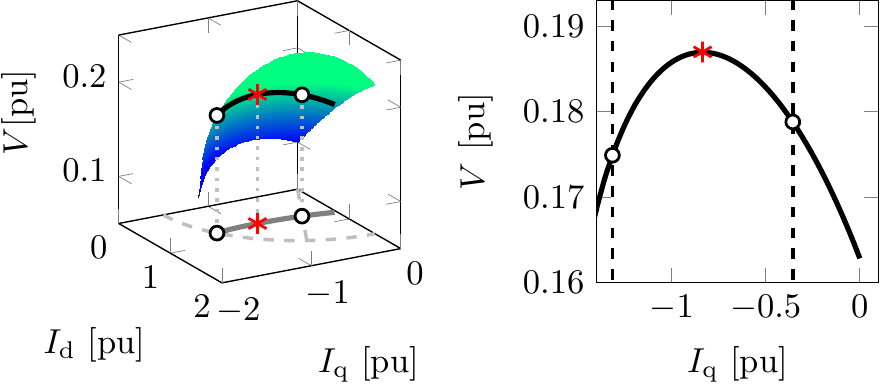}
    \caption{Dimensionality reduction based on $B_p$. Top ($S_2$): $V_g=0.5$ pu, $Z=0.1$ pu, $R/X=2$, $P_{\rm max}=0.436$ pu; bottom ($S_3$): $V_g=0.1$ pu, $Z=0.1$ pu, $R/X=2$, $P_{\rm max}=0.126$ pu.  }
    \label{3Dto2D_S23}
\end{figure}

This reveals that there is a unique maximizer $I_{\rm q}^\star$ of $V$ over a specific interval $[\underline{I_{\rm q}},\overline{I_{\rm q}}]$, which corresponds to $S_3$ if C2 holds or $S_2$ otherwise. The upper bound  $\overline{I_{\rm q}}$ is given  in (\ref{Iqmax}), which in geometry corresponds to the unique intersection point between ${L}_1$ and $B_p$ in the fourth quadrant. It should be clarified that $\overline{I_{\rm q}}$ is somewhat conservative, i.e., $V$ is expected to  decrease over a wider range (see examples later in Fig. \ref{3Dto2D_S23}). The lower bound $\underline{I_{\rm q}}$ is determined by CC or SSC.  If the grid is weak and/or the voltage dip is severe [so that the condition in (\ref{C3}) does not hold], $\underline{I_{\rm q}}$ is determined by SSC; otherwise, it is determined by CC. 
Therefore, searching for $S_2$ or $S_3$ can be achieved by searching for $I_{\rm q}^\star$ over $[\underline{I_{\rm q}},\overline{I_{\rm q}}]$ if C1 does not hold.

Fig. \ref{3Dto2D_S23} gives two illustrative examples of dimensionality reduction associated with $B_p$. In the first (top) case  where neither C1 nor C2 holds, $V$ strictly decreases within $[-1.318,-0.374]$ pu and has a unique local maximizer $I_{\rm q}^\star=-0.374$ pu, which exactly corresponds to $S_2$ computed by (\ref{OPTS2}); in the second  (bottom) case where C2 holds, $V$  strictly increases within $[-1.316,-0.836]$ pu and strictly decreases within $[-0.836, -0.356]$ pu. It has a unique local maximizer, $I_{\rm q}^\star=-0.836$ pu, which exactly corresponds to $S_3$  computed by (\ref{OPTS3}). 

\section{Optimum Seeking}
Thanks to the dimensionality reduction, solving the ODVS problem (\ref{ODVS}) can be reduced to a 1-D search no matter the value of $P_{\rm max}$. More precisely, one can search for the maximizer $x^\star$ of a 1-D \emph{unimodal} function $V(x)$ over a specific interval $[\underline{x},\overline{x}]$, where $x$ denotes either $\phi$ or $I_{\rm q}$. Note that we have only shown the existence of such a 1-D functional relationship, but its explicit expression is, in fact, not available due to the lack of grid parameters. So, an immediate question of interest is: \emph{how to search for $x^\star$ efficiently without the need to know the explicit expression of $V(x)$? }

Perturb-and-observe is considered an effective and easy-to-implement method for 1-D search, provided the objective function value is available. 
Fortunately, the PoC voltage, i.e., the objective function,  is usually measurable in practice. This enables us to develop a P\&O-based OS algorithm, of which the updated rule at the $k$th step is given as
\begin{align}
    x_{k+1}&\leftarrow\left[x_k+\alpha_k\cdot d_k\right]_{\underline{x}}^{\overline{x}}\label{PnO1}\\
    d_{k+1}&\leftarrow{\rm sgn}\big(V(x_k)-V(x_{k-1})\big)\cdot{\rm sgn}\big(d_{k}\big)\label{PnO2}
\end{align}
where $\alpha_k$ is the perturbation stepsize; $d_k$ is the search direction; $[\cdot]_{\underline{x}}^{\overline{x}}$ denotes the projection operator onto the set $[\underline{x},\overline{x}]$; ${\rm sgn}(u)=1$ for $u\geq0$ and ${\rm sgn}(u)=-1$ for $u<0$. 

Of great importance will be the design of stepsize rule $\{\alpha_k\}$, which dictates the convergence performance. Conventionally, the stepsize can be fixed, but its selection is very tricky---a large stepsize could lead to significant steady-state oscillations, whereas a small stepsize could slow down the OS. 
In light of this, we provide the following rigorous criteria and practical examples for stepsize rule design to enhance  convergence.

\emph{Theorem 1:} The sequence $\{x_k\}$ generated by  (\ref{PnO1})--(\ref{PnO2}) will converge to the optimum $x^\star$, if $0<\alpha_k<\infty$ for all $k$, and
\begin{align}\label{ConvergenceCon}
     \lim_{k\rightarrow\infty} \alpha_k=0,\hspace{3mm}    \sum_{k=0}^{\infty}\alpha_k=\infty.
    \end{align}

This suggests that the sequence $\{\alpha_k\}$ should be \emph{diminishing} and \emph{non-summable}, as $k\rightarrow\infty$. The property of ``\emph{diminishing}" guarantees zero steady-state oscillation around the limit point of $\{x_k\}$, while the property of ``\emph{non-summable}" guarantees the optimality of the limit point.
There are indeed numerous available stepsize rules meeting the above criteria. Below, we provide an option based on the ``$p$-series.”

\emph{Corollary 1:} Suppose  $0<p\leq1$ and $\lambda>0$. By taking the stepsize rule:
\begin{align}
    \alpha_k=\frac{\lambda}{k^p}
\end{align}
the sequence $\{x_k\}$ generated by  (\ref{PnO1})--(\ref{PnO2}) will converge to the optimum $x^\star$.

$\lambda$ is a scale factor that  dictates the \emph{average} convergence performance throughout the whole process. A larger $\lambda$ can accelerate the convergence but may, in turn, cause oscillations. $p$ determines the diminishing rate of stepsize and thus dictates the convergence performance at different stages of the process. A larger $p$ can alleviate large oscillations but possibly slow down the convergence. So, good convergence dynamics should be established on a suitable combination of $\lambda$ and $p$. 

Keep in mind that the convergence performance  should definitely  be related to the unknown grid parameters, in the sense that these parameters dictate the shape of $V(x)$ and the value of $[\underline{x},\overline{x}]$; the same stepsize rule may work out differently under different grid conditions. Therefore, thorough offline trials and analyses under different grid conditions are advocated to ensure the robustness of the stepsize rule.

\section{Implementation}
So far in this paper, we have elaborated in theory on how to achieve the dimensionality reduction and how to perform the OS algorithm thereafter. With those  fundamentals in place, in this section, we will present the implementation method of the model-free ODVS, which will be embedded within a single-stage PV system.
\subsection{Perturbed Variable and OS Mode}
In Section IV, we base the OS on the prior knowledge of the perturbed variable ($\phi$ or $I_{\rm q}$); in other words, the valid stage of optimum ($S_1$ or $S_2/S_3$) should have been known before OS. It should be clarified that there is no need to further distinguish between $S_2$ and $S_3$ as they share the same perturbed variable and the corresponding OS algorithm. Accordingly, there will be two OS modes: OS-a for $S_1$ and OS-b for $S_2/S_3$. 

Recall in Section II that C1 gives a theoretical criterion for OS mode selection. Unfortunately, it cannot be straightforwardly used since the grid parameters are unavailable. However, the rationale  behind C1---whether or not the IBR can provide adequate active power in favor of $S_1$---is clear and useful. Inspired by this, we propose the following strategy to handle the mode selection issue. \emph{The IBR operates with OS-a by default; if the system cannot afford the required active power corresponding to $S_1$, then switch the OS to OS-b.} 

Whether $S_1$ is affordable can be reflected by the dc voltage behavior  of the IBR, provided the primary dc-side source injects the maximum available power; more precisely, the power mismatch between the dc and ac sides will induce significant dc voltage drops once the dc-side source cannot offer the required power corresponding to $S_1$. So, an easy-to-implement criterion for mode switching is thus given by
\begin{align}\label{dcvoltth}
    \widetilde{V}_{\rm dc}\leq\rho V_{\rm dc}^{\ast}
\end{align}
where ${V}_{\rm dc}^{\ast}$ is the dc voltage reference of the IBR  immediately before the PoC voltage drops, which is held constant over the whole period of DVS; $\widetilde{V}_{\rm dc}$ is the notch-filtered dc voltage; the coefficient $\rho$ ($0<\rho<1$) determines the  threshold.
\subsection{Enforcement of $B_c$ and $B_p$}
A prerequisite for the dimensionality reduction (as well as the OS) is that the IBR shall always operate along the maximum current boundary $B_c$ (for OS-a) or the maximum power boundary $B_p$ (for OS-b). The former can be achieved in a feedforward manner, as shown in Fig. \ref{DVScontroller}.
The latter can be achieved by combining the dc-side maximum power control and the dc voltage control. For instance, consider a single-stage PV system in this work. The dc voltage of the PV inverter is controlled (by regulating $I_{\rm d}$) to track the dc voltage reference yielding the maximum power point of PV panels. 

\subsection{Feasibility of Perturbed Variable}
Recall in Section IV that $x_k$ should be restricted within a specific interval $[{\underline{x}},{\overline{x}}]$ at each iteration, so that the convergence of OS can be guaranteed. Unfortunately, the lack of grid parameters makes it impossible to directly compute $\underline{x}$ and $\overline{x}$, as they are related to the grid parameters. So instead, we will  leverage the physical essence of such bounds to avoid the violation.

\emph{OS-a:}
According to the analysis in Section III-B, $[\underline{\phi},\overline{\phi}]$ is determined by SSC and MAPC. $[\underline{\phi_s},\overline{\phi_s}]$ corresponds to SSC, which is not violated provided the IBR maintains synchronism with the grid; we will discuss how to physically achieve this shortly. Theoretically, it is hard to guarantee that OS-a does not violate MAPC due to the lack of grid parameters. However, the violation of MAPC is very unlikely to occur in practice if C1 holds.\footnote{This is because if the violation of MAPC occurs, $P_b\leq P_{\rm max}$ and $g_p(\phi)>0$ for some $\phi\in[\underline{\phi_s},\overline{\phi_s}]$ should be satisfied, simultaneously. Let us take the examples in Fig. \ref{3Dto2D_S1} again. Accordingly, $P_{\rm max}$ should exactly fall within $[0.7375,0.8012]$ and $[0.5367, 0.5525]$ pu, respectively, of which the probabilities are very low.
Note that this is only a necessary condition for the violation of MAPC. That is, even though $P_{\rm max}$ exactly falls within that specific interval, it does not mean that the violation will definitely happen; it should also depend on the design of the OS controller. For example, a proper stepsize rule or initialization may help avoid such violation.}  Even though the violation of MAPC does occur in practice, according to Section V-A, the OS will be switched from OS-a to OS-b once a significant dc voltage is detected. This will cause the system to reach a near-optimal solution, instead of becoming unstable. 

\begin{figure}[t]
    \centering
    \includegraphics[width=3.5in]{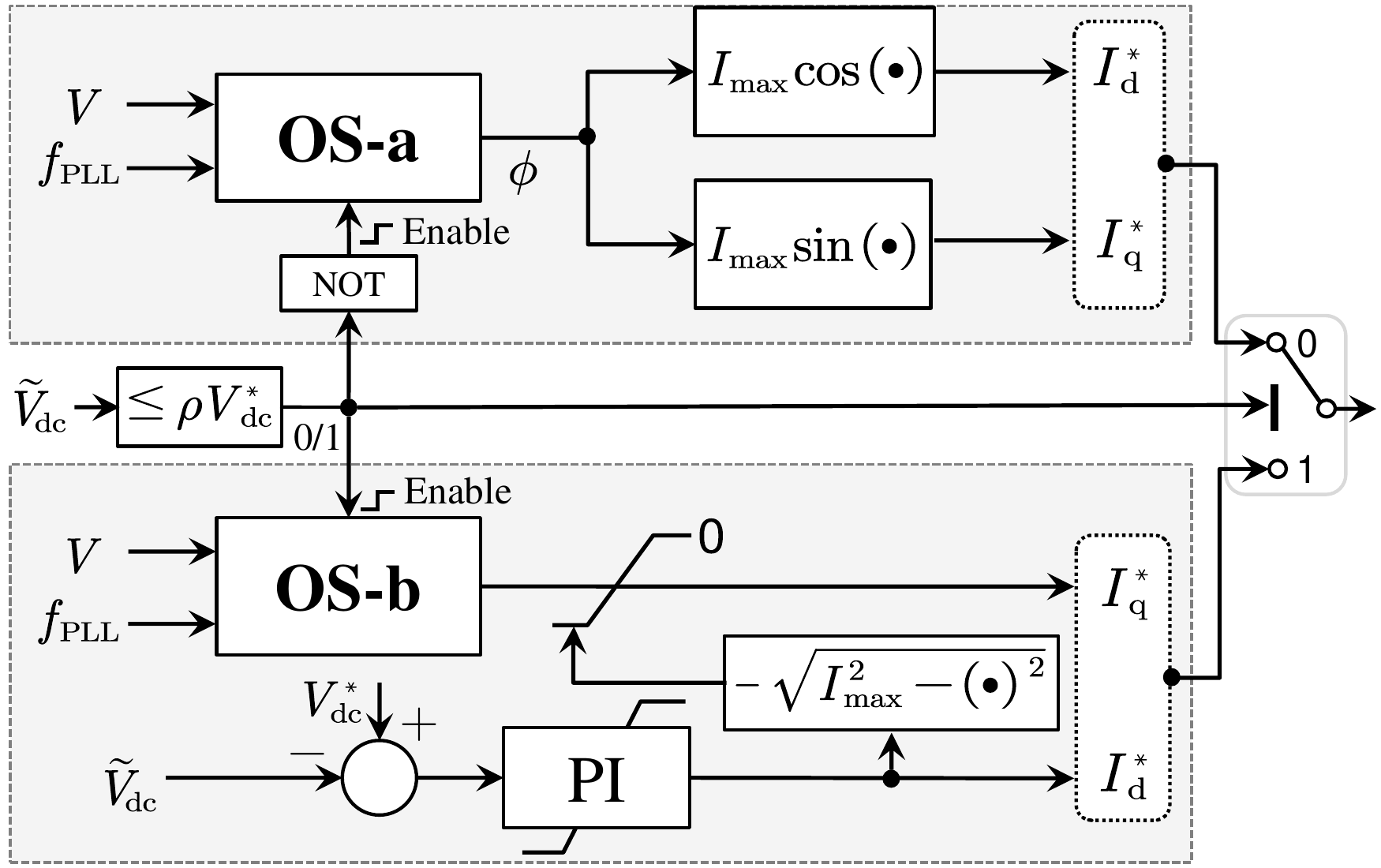}
    \caption{Model-free DVS controller design. }
    \label{DVScontroller}
\end{figure}

\emph{OS-b:} According to the analysis in Section III-C,  $\underline{I_{\rm q}}$ corresponds to either CC or SSC. This implies that it is not violated, provided  neither of those two constraints is violated. As depicted in Fig. \ref{DVScontroller}, two additional current limiters can accomplish the current limiting. As mentioned above, SSC is not violated, provided the IBR can remain synchronized with the grid.
In contrast,  $\overline{I_{\rm q}}$ does not directly correspond to any physical constraint of an IBR; hence, theoretically, it is unlikely to know whether such bound is violated or not. But as mentioned earlier, $\overline{I_{\rm q}}$ is rather conservative; thus, in reality, a moderate violation does not undermine the convergence of OS\footnote{It is expected that the monotonicity of objective function (strictly decreasing) is preserved for $\overline{I_{\rm q}}\leq I_{\rm q}\leq 0$, though it is hard to rigorously prove this result; see the numerical examples in Fig. \ref{3Dto2D_S23}.}. Since the optimum must be in the fourth quadrant, a conservative but effective upper bound, $I_{\rm q}\leq0$, can be imposed. Besides, as a remedy, we can carefully initialize the OS to prevent it from reaching the upper bound\footnote{If the R/X ratio of the main grid is roughly known, one can obtain a crude estimate of $[\underline{I_{\rm q}},\overline{I_{\rm q}}]$ and then initialize the OS by letting $x_0=(\underline{I_{\rm q}}+\overline{I_{\rm q}})/2$; otherwise, just simply, let $x_0=-I_{\rm max}/2$.}. 

\emph{Synchronization Stability:} The remaining issue is to guarantee that the IBR maintains synchronism with the grid during the OS. As known, when LoS occurs, the frequency of the PoC voltage calculated by the PLL ($f_{PLL}$) significantly deviates from the nominal frequency $f_N$ \cite{GO2014}, which can be considered a sign of the violation of SSC. 
LoS usually results in significantly distorted PoC voltage, under which the OS may fail. The instability is even exacerbated due to the interaction between the OS and the system's transient  dynamics immediately after the voltage drops. To avoid this, we propose the following ``freezing strategy.” Once the PLL frequency deviation exceeds a given tolerance $\Delta f$, i.e., 
\begin{align}\label{PLLfreq}
    |f_{PLL}-f_N|\geq\Delta f,
\end{align}
the OS will be frozen with a fixed output $\phi=-45^\circ$ (for OS-a) or $I_{\rm q}=-I_{\rm max}/4$ (for OS-b). When the PLL frequency returns within the deadband $ |f_{PLL}-f_N|<\Delta f$, the OS will be unfrozen.
\subsection{Implementation Within A Single-Stage PV System}
Based on the above discussions, the overall controller design of model-free DVS  is shown in Fig. \ref{DVScontroller}, which is embedded within a single-stage PV system. We do not modify  the control scheme of the PV system under normal operating conditions. The whole procedure of DVS in response to a voltage dip is further detailed  below. 
\begin{itemize}
    \item During normal operating conditions, the single-stage PV system adopts the MPPT strategy and operates with a unity power factor (steady-state reactive power support from the PV inverter is also allowed if necessary). 
    \item If the positive-sequence PoC voltage drops below a given threshold (e.g., $0.9$ pu), the MPPT control is immediately frozen, and the generated dc voltage reference is held constant afterwards\footnote{Here, an implicit assumption is that the solar irradiance does not significantly change during the DVS. This is widely believed to hold in practice since the duration of DVS is typically very short (several tens of milliseconds to several seconds).}. Meanwhile, the DVS is triggered.
    \item During the DVS, OS-a is first carried out by default, i.e., $\phi$ is perturbed to seek the maximum voltage support based on the iterative algorithm (\ref{PnO1})--(\ref{PnO2}). 
    \item If the condition (\ref{dcvoltth}) is satisfied,  the control mode will be switched from OS-a to OS-b, where $I_{\rm q}$ becomes the perturbed variable; otherwise, OS-a is kept activated. 
    \item The DVS controller generates the (positive-sequence) current references $I_{\rm d}^{\ast}$ and $I_{\rm q}^{\ast}$ for the inner current control of the inverter.
\end{itemize}
\begin{table}[t]
\small
\centering
\caption{System Parameters}\label{mainparameters}
\renewcommand\arraystretch{0.95}
\begin{tabular}{ m{1.3in} m{1.7in}}
        \hline\hline
Description&Value\\
         \hline
Power rating of PV& 250 kW (1 pu) \\
Nominal ac voltage& 250V/25kV (1 pu)\\
Nominal frequency & 60 Hz\\
Pre- and post-fault SCRs& 20, 10\\
R/X ratio&2\\
Maximum current limit& 1.5 pu\\
Sampling rate of OS & 30 Hz\\
$\rho$ & 0.95\\
$\Delta f$& 0.3 Hz\\
$x_0$ & $-45^\circ$ (OS-a), $-0.75$ pu (OS-b)\\
$d_0$ & $-1$ (OS-a/b)\\
$\lambda$ & 15 (OS-a), 0.2(OS-b)\\
$p$ & 1 (OS-a/b)\\
 \hline\hline
        \end{tabular}
\end{table}
\section{Simulation Results}
We test the model-free ODVS control with a grid-connected PV system via dynamic simulations in the MATLAB/Simulink R2021a environment. The main specifications of the test system are given in Table \ref{mainparameters}. Four test cases (A,B,C,D) are considered to examine the DVS control performance under different operating conditions. We compare the model-free ODVS method with the model-based ODVS  method \cite{YG2021} and the widely-used droop-based DVS strategy \cite{AS2019}. In the model-based ODVS, the accurate grid parameters are assumed to be available at the very beginning of DVS; hence, the ODVS can be performed  without any delay\footnote{Here, the implementation method of the model-based ODVS is the same as in \cite{YG2021}, except that it no longer requires a three-cycle delay to accomplish grid estimation. Though it is somewhat unrealistic for real-world applications, it can be used as the benchmark to validate the effectiveness of the model-free ODVS.}. In the droop-based DVS strategy, the reactive current injection is controlled as per the piece-wise linear rule \cite{AS2019}:
\begin{align}
\nonumber I_{\rm q}=\left\{\hspace{-2mm}
   \begin{array}{cl}
    -{I}_{\rm max},& {\rm if}\,\,V\leq0.5\\[2mm]
    \dfrac{0.9-V}{0.5-0.9}\cdot{I}_{\rm max},  & {\rm if}\,\,0.5<V<0.9\\[3mm]
    0,& {\rm if}\,\,V\geq0.9.
   \end{array}
   \right.
\end{align}
The active current reference is generated by the dc voltage PI controller (same as in OS-b) but note that the reactive current is prioritized (different with OS-b) in the droop-based strategy.
\subsection{Search for $S_1$}
\begin{figure}[t]
    \centering
    \includegraphics[width=3.5in]{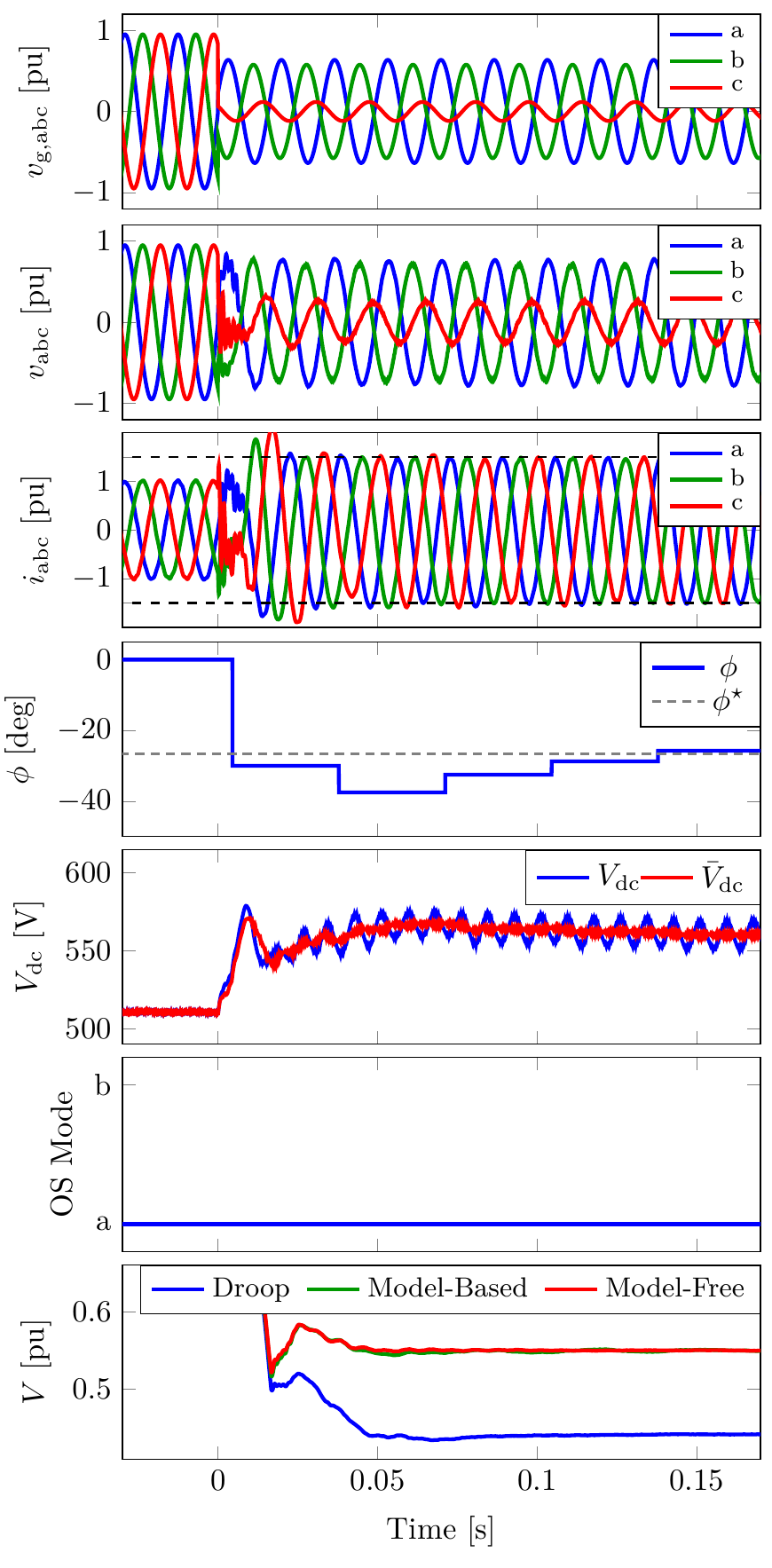}
    \caption{Simulation results---Case A. From top to bottom: 1) three-phase grid voltage, 2) three-phase PoC voltage, 3) three-phase inverter current injection,  4) power factor angle, 5) dc voltage, 6) OS mode selection, and 7) positive-sequence PoC voltage under different control strategies. }
    \label{figs_caseA}
\end{figure}
\begin{figure}[t]
    \centering
    \includegraphics[width=3.5in]{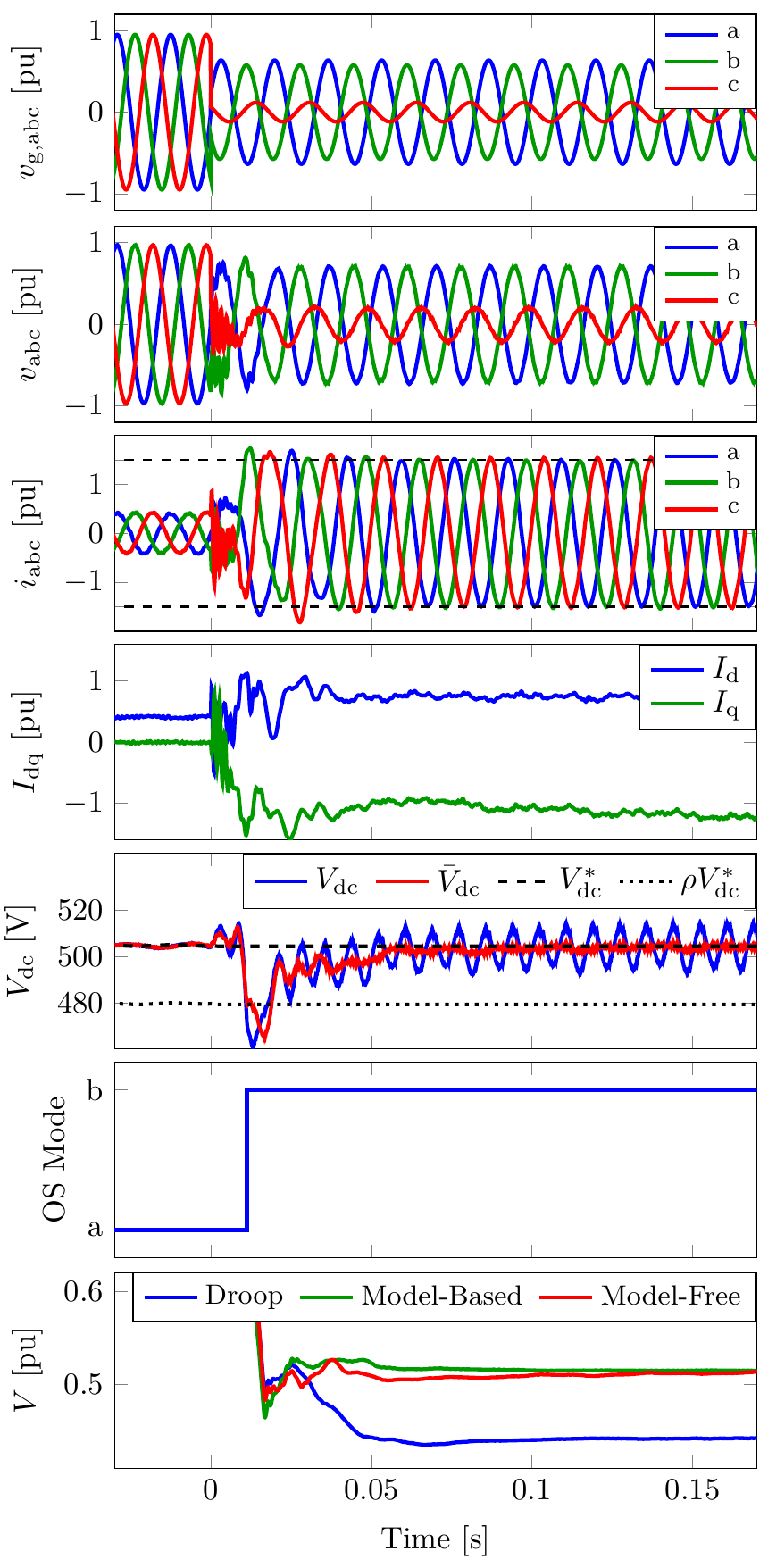}
    \caption{Simulation results---Case B. From top to bottom: 1) three-phase grid voltage, 2) three-phase PoC voltage, 3) three-phase inverter current injection, 4) active and reactive currents,  5) dc voltage, 6) OS mode selection, and 7) PoC voltage under different control strategies.}
    \label{Simulation_Case B}
\end{figure}
In Case A, an unbalance grid voltage dip occurs at $t=0$s, where the positive-sequence and negative-sequence components are $0.4\angle0^\circ$pu and $0.3\angle50^\circ$pu, respectively. The solar irradiance is $1000\rm W/m^2$, and the cell temperature is $25^\circ\rm C$. In this context, the IBR shall operate with $S_1$ to  achieve the maximum DVS. It can be seen from Fig. \ref{figs_caseA} that the model-free ODVS control is triggered immediately after the voltage dip happens. It constantly operates with OS-a because the PV generation can afford the required power corresponding to $S_1$ ($184$ kW). Due to the self-protection capability of a single-stage PV generation system, the dc voltage is boosted to a new equilibrium ($\approx 560$V); the mode switching criterion in (\ref{dcvoltth}) is never met. The perturbed variable $\phi$ of OS quickly converges to the optimum $\phi^\star={\rm atan2}(-X,R)=-26.57^\circ$ within a few iterations ($\leq5$ iterations), and therefore, the PoC voltage is recovered rapidly to $0.55$pu---the maximum DVS. It achieves almost the same DVS performance as the model-based ODVS, where the optimal active and reactive current set-points are directly given and outperforms the droop-based strategy, by which the PoC voltage can only be  recovered to $0.44$pu. 
\subsection{Search for $S_2$}
The simulation settings in Case B are the same as in Case A, except for the solar irradiance, which is set as $\rm 400W/m^2$ here. So, the IBR shall operate with $S_2$ to achieve the maximum DVS. Fig. \ref{Simulation_Case B} showcases the simulation results. After the DVS control is triggered, OS-a is first carried out by default. It is observed that the dc voltage rapidly drops below the threshold $95\%V^\ast_{\rm dc}$ ($\approx480$V) since the PV generation cannot provide sufficient power to support $S_1$ anymore under such a solar irradiance condition, after which the OS switches from OS-a to OS-b (at $t=0.0112$s). Then, the dc voltage is recovered quickly to the reference value ($\approx505$V) within $30$ms. Meanwhile, the system rapidly converges to the optimum S2 within $40\sim50$ms through the OS, and correspondingly, the PoC voltage is recovered to $0.515$pu---the maximum DVS. The model-free ODVS achieves the similar DVS performance as the model-based ODVS (both of them converge to the optimum). However, the dynamics are slightly different, and it performs better than the droop-based strategy, by which the PoC voltage is recovered only to $0.44$pu. 
\subsection{Search for $S_3$}
\begin{figure}[t]
    \centering
    \includegraphics[width=3.5in]{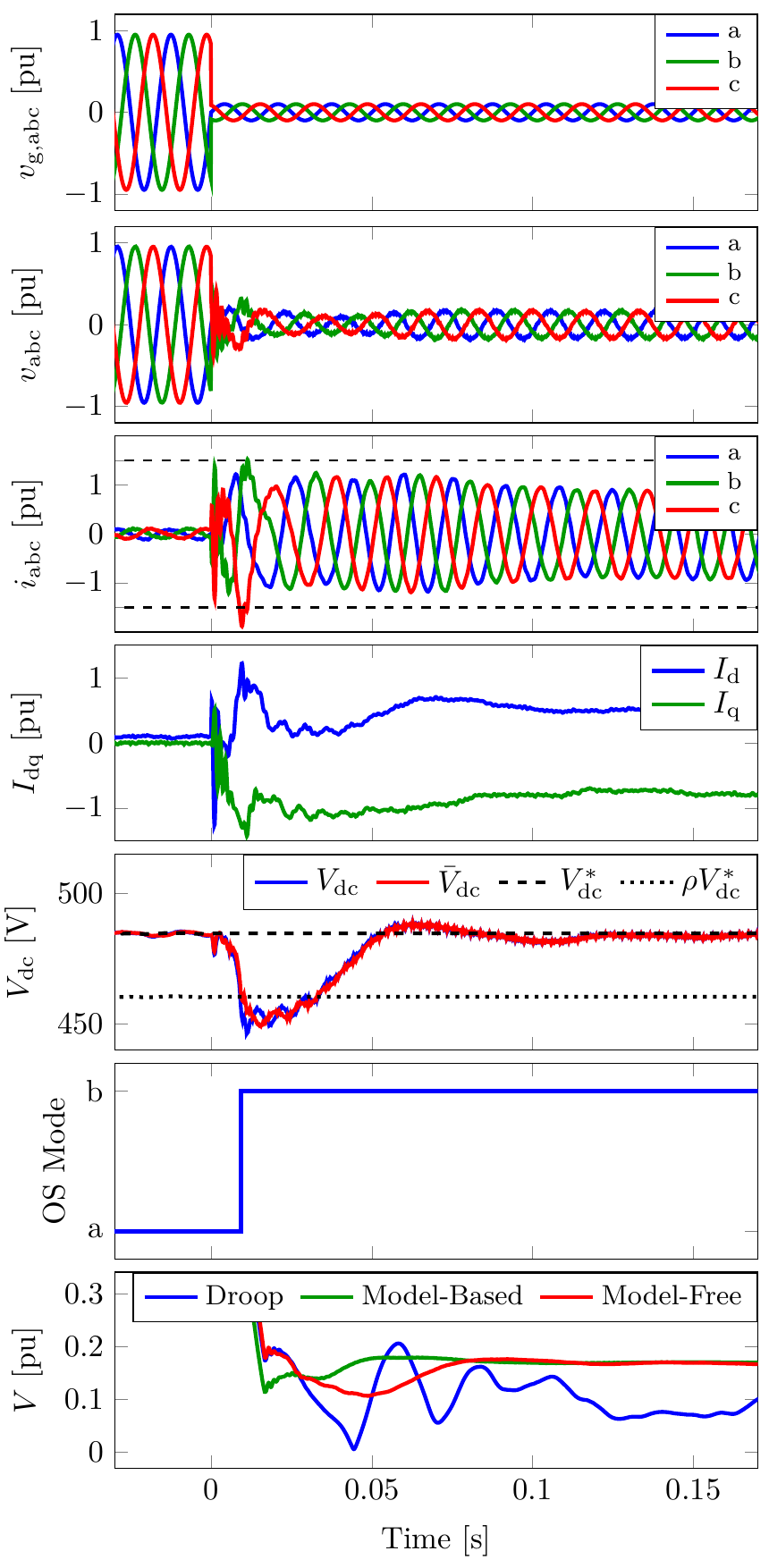}
    \caption{Simulation results---Case C. From top to bottom: 1) three-phase grid voltage, 2) three-phase PoC voltage, 3) three-phase IBR current injection, 4) active and reactive currents,  5) dc voltage, 6) OS mode selection, and 7) PoC voltage under different control strategies.}
    \label{Simulation_Case C}
\end{figure}
In Case C, a balanced voltage dip ($V_g=0.1\angle0^\circ$pu) is simulated at $t=0$s; the solar irradiance is set as $100\text{W}/\text{m}^2$. So, the IBR shall operate with $S_3$ for the maximum DVS. The simulation results are depicted in Fig. \ref{Simulation_Case C}. Similar to Case B, OS-a is first enabled by default, then switched to OS-b as the dc voltage drops below $95\%V_{\rm dc}^\ast$ ($\approx460$V). The model-free ODVS drives the IBR system towards the optimum $S_3$, where the PoC voltage is recovered to $0.169$pu---the maximum DVS. The model-free and model-based ODVS methods ultimately achieve a similar DVS performance after the system settles down. In contrast, the droop-based strategy results in significantly distorted PoC voltage because it drives the system towards an infeasible operating point that potentially violates SSC and thus induces the LoS.
\subsection{Test of ``Freezing Strategy"}
\begin{figure}[t]
    \centering
    \includegraphics[width=3.5in]{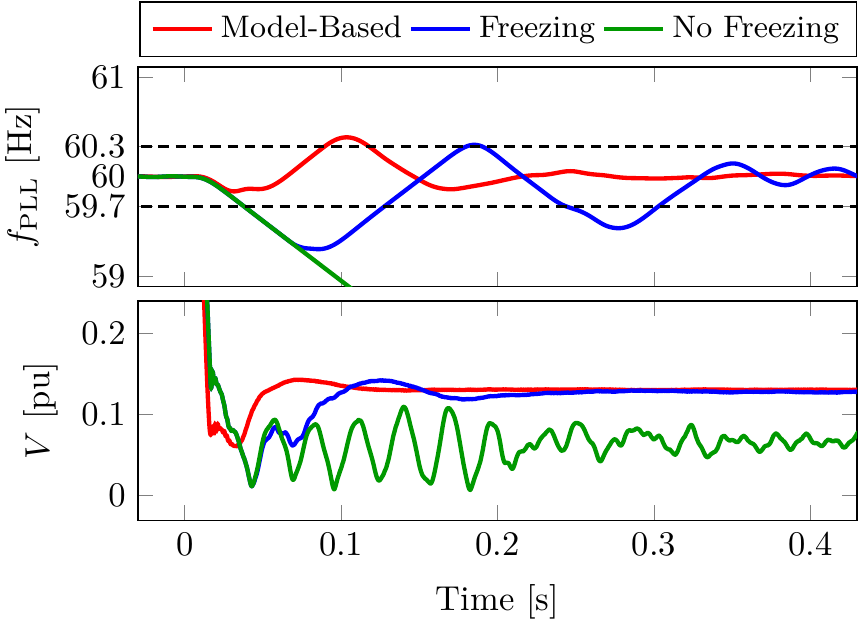}
    \caption{Simulation results---Case D. PLL frequency (top) and positive-sequence PoC voltage (bottom) under different control strategies.}
    \label{Simulation_Case D}
\end{figure}

Many grid codes require the connectivity of IBRs for a certain length of time (e.g., 150 ms \cite{GB19964}) even when the voltage drops below 10\% of the nominal voltage, whereby LoS is very likely to happen. So, in case D, we consider a very severe voltage dip by setting $V_g=0.05$ pu; other simulation settings are the same as in Case C. As shown in Fig. \ref{Simulation_Case D}, without the freezing strategy, the system loses synchronism quickly with the main grid, which will result in disconnection of the IBR in practice. In contrast, the model-free ODVS with the freezing strategy promptly stabilizes the system, 
reflected by the fact that the PLL frequency is quickly corrected close to the nominal value. Consequently, the system not only avoids LoS but also converges to the optimal operating point corresponding to the maximum DVS (similar to the model-based ODVS). This demonstrates the necessity of such a freezing strategy, especially under a severe voltage dip. 
\section{Conclusions}
This paper proposes a model-free ODVS control method for maximizing the DVS  provided by IBRs without relying on the knowledge of grid parameters. We first determine  the potential binding constraints of the ODVS problem, CC, and MAPC. Then, we develop a P\&O-based OS algorithm to search for the optimum along the boundary of CC and MAPC. The algorithm is provably convergent to the optimum provided the stepsize sequence is diminishing and non-summable.
It is demonstrated that the proposed model-free ODVS responds to voltage dips promptly and achieves similar DVS performance as the model-based ODVS does (the optimality is guaranteed). Moreover, it has several other advantages, such as inherent robustness against model mismatch, low computational burden, and dynamic  responsiveness.

\appendices

\section{Proof of Theorem 1}
\subsection{Proof of Theorem 1}
\emph{Proof of OS-a:} We first show that there does \emph{not} exist a positive integer $m$, such that for all $k\geq m$, $\phi_k>\phi^\star$ or $\phi_k<\phi^\star$. Assume the contrary, that is, there exists some $m$, such that,
\begin{align*}
   (\text{Case a}):\hspace{3mm} \phi_k&<\phi^\star,\hspace{3mm} \forall k\geq m \hspace{3mm} {\rm or}\\
    (\text{Case b}):\hspace{3mm} \phi_k&>\phi^\star,\hspace{3mm} \forall k\geq m.
\end{align*}

Accordingly, due to the property of function $V(\phi)$ and the rule of OS algorithm, there must exist some $i\geq m$, such that,
\begin{align*}
    &{V}(\phi_i)<{V}(\phi_{i+1})<{V}(\phi_{i+2})<\cdots\\
   & d_i=d_{i+1}=d_{i+2}=\cdots=\left\{\begin{matrix}1, \text{Case a}\\-1, \text{Case b}\end{matrix}\right.
\end{align*}
and thus, 
$$
\phi_{i+1}=\big[\phi_i+\alpha_id_i\big]^{\overline{\phi}}_{\underline{\phi}}=\phi_i+\alpha_id_i.
$$
Since the stepsize sequence is non-summable, it follows that,
\begin{align*}
    &\lim_{k\rightarrow\infty} \phi_k=\phi_i+\lim_{k\rightarrow\infty}\sum_{j=i}^{k-1}\alpha_j=\infty>\phi^\star,\,\,\,(\text{Case a})\\
    &\lim_{k\rightarrow\infty} \phi_k=\phi_i-\lim_{k\rightarrow\infty}\sum_{j=i}^{k-1}\alpha_j=-\infty<\phi^\star,\,\,\,(\text{Case b})
\end{align*}
which contradicts our earlier assumption. 

So, this means that $\phi_k$ will always jump around the optimum $\phi^\star$ as $k\rightarrow\infty$; so, there exists some positive integer $m$, such that,
\begin{align*}
    \phi_{m-1}&<\varphi^\star\leq \phi_m,\,\,\,\text{or}\\
     \phi_{m-1}&>\varphi^\star\geq \phi_m,
\end{align*}
which yields,
\begin{align*}
    \left\|\phi_m-\phi^\star\right\|<\left\|\phi_{m}-\phi_{m-1}\right\|=\alpha_{m-1}.
\end{align*}
Then, it follows that
\begin{align*}
    \left\|\phi_{m+1}-\phi^\star\right\|&=\left\|\big[\phi_m+\alpha_md_m\big]_{\underline{\phi}}^{\overline{\phi}}-\phi^\star\right\|\\
    &\leq\left\|\phi_m+\alpha_md_m-\phi^\star\right\|\\
    &\leq\left\|\phi_m-\phi^\star\right\|+\left\|\alpha_m\right\|\\
    &<\left\|\alpha_{m-1}\right\|+\left\|\alpha_m\right\|.
\end{align*}

Since the iteration will always switch between $\phi_k\leq\phi^\star$ and $\phi_k\geq\phi^\star$ as $k\rightarrow\infty$. Define a \emph{cycle} as the procedure that starts from the $m$th iteration where $\phi_{m-1}<\phi^\star\leq\phi_m$ and ends up with the $n$th iteration where $n$ is the minimum number satisfying $\phi_{n}<\phi^\star\leq \phi_{n+1}$; let $\mathcal{K}(c):=\{m,m+1,m+2,...,n\}$ be the set of indexes associated with the $c$th cycle and 
$D(c):=\max_{k\in \mathcal{K}(c)}\left\|\phi_k-\phi^\star\right\|.$ Then, we have
$$
0\leq D(c)<\left\|\alpha_{m-1}\right\|+\left\|\alpha_{m}\right\|
$$

Given that the stepsize sequence is diminishing, we have (by the Squeeze Theorem)
\begin{align*}
    \lim_{c\rightarrow\infty} D(c)=0
\end{align*}
which indicates that 
\begin{align*}
\lim_{k\rightarrow\infty} \left\|\phi_k-\phi^\star\right\|=0.
\end{align*}
This completes the proof of OS-a. \qed

\emph{Proof of OS-b:} If $S_3$ is valid, the proof is similar to the proof of OS-a, which is thus omitted for brevity. Let us consider the case where $S_2$ is valid, that is, ${V}(I_{\rm q})$ is strictly decreasing over $[\underline{I_{\rm q}},\overline{I_{\rm q}}]$. 

We first show that there must exist some  $k$ such that $I_{{\rm q},k}=\underline{I_{\rm q}}=I_{\rm q}^\star$. Similarly, assume the contrary, that is, $I_{{\rm q},k}>\underline{I_{\rm q}}$ for any $k$. Then, as per the  OS algorithm and the monotonicity of ${V}(I_{\rm q})$, we have for any $k\geq1$
\begin{align*}
I_{{\rm q},k+1}&=\Big[I_{{\rm q},k}+\alpha_{k}d_{k}\Big]_{\underline{I_{\rm q}}}^{\overline{I_{\rm q}}}=I_{{\rm q},k}-\alpha_{k}.
\end{align*}
Given that the stepsize sequence is non-summable, we have
\begin{align*}
    \lim_{k\rightarrow\infty} I_{{\rm q},k}&=I_{{\rm q},1}-\lim_{k\rightarrow\infty} \big(\alpha_1+\cdots+\alpha_{k-1}\big)=-\infty<\underline{I_{\rm q}}
\end{align*}
which contradicts our earlier assumption. 

So, there must exist some $k$, such that $I_{{\rm q},k}=\underline{I_{\rm q}}=I_{\rm q}^\star$.
Then, due to the iteration rule of OS, 
it follows that,
\begin{align*}
 I_{{\rm q},m}=I_{{\rm q},k}=\underline{I_{\rm q}}=I_{\rm q}^\star,\,\,\,\forall m\geq k.
\end{align*}
In other words, once the OS reaches the optimum, it will be stuck there. This completes the proof of OS-b. \qed

\section{Proof of Corollary 1}
It is not hard to see that
\begin{align*}
    0<\frac{\lambda}{k^p}&<\infty, \,\, \text{for all}\,\,\,k\\
     \lim_{k\rightarrow\infty} \frac{\lambda}{k^p}&=0, \\
     \sum_{k=0}^{\infty}\frac{\lambda}{k^p}&=\infty,
\end{align*}
for $0<p\leq 1$ and $\lambda>0$. Then, according to Theorem 1, the convergence of OS is guaranteed. \qed
\bibliographystyle{IEEEtran}
\bibliography{references}

\end{document}